# Analytical Model of Fast Ion Behavior in Current Hole Tokamak Plasmas

K. Schoepf[a], V. Yavorskij[a,b], V. Goloborod'ko[a,b], P. Neururer[a]

[a]*Institute for Theoretical Physics, University of Innsbruck, Austria, Association EURATOM-OEAW*
[b]*Kiev Institute for Nuclear Research, Ukrainian Academy of Sciences, Kiev, Ukraine*

**Introduction**

As demonstrated by theory and experiments, e.g. [1-4], optimization of the magnetic field topology in tokamaks by specifying favorable current profiles can improve the plasma confinement at reduced inductive current. One such promising approach is a so-called current hole scenario featuring a central plasma region with no poloidal field due (near) zero toroidal current density there. The establishment of a current hole (CH) occurs typically with the formation of internal transport barriers suppressing the bulk plasma transport within the current hole and allowing for stable plasma equilibrium over remarkably long periods [5-8]. Though a CH regime is recognized to provide better detention of the bulk plasma, it may negatively act on the confinement of fast ions [9-12] such as fusion products and neutral-beam injected (NBI) ions. Since, however, the transport properties of these energetic particles determine the heating profiles and the power loading upon the first wall, and therefore are of crucial importance in a fusion reactor, we examine here analytically the CH effects on the fast ion behavior in a tokamak. For that we employ a simplified model based on an analytical approximation of the poloidal flux function [10] allowing for a complete characterization of possible orbit topologies. In the constants-of-motion space we determine the confinement domains for the different types of ion orbits, calculate the CH induced alterations of the fast ion transport and derive the distribution of NBI ions for a specific JET CH plasma scenario.

**Current profile model**

The poloidal flux $\Psi(r)$ with $r$ denoting the equatorial distance of a considered flux surface from the magnetic axis is capable of describing completely the impact of a toroidal current hole on particle motion and orbital trajectories. Normalizing with respect to the minor plasma radius $a$ we introduce the normalized flux surface radius $x = r/a$ and adopt, for CH tokamak equilibria, the simple linear model from Ref. [10],

$$\Psi(x, x_*) \approx (x - x_*)\Psi'_a \Theta(x - x_*) \quad , \qquad (1)$$

in which $x_* = r_*/a$ is the effective hole size of the poloidal flux and thus constitutes a measure for the toroidal current hole, $\Theta$ represents the Heaviside step function and the derivative at the plasma edge is $\Psi'_a = 2aA\gamma I/c$ with the aspect ratio $A$, the total plasma current $I$, the speed of light in free space, $c$, and a plasma shape factor $\gamma$ [13]. To fit to real experiments, the effective hole size $x_*$ is suggested to be approximated via minimization of the functional

$$W(x_*, x_\phi) = \int_0^1 dx \left\{ \Psi(x, x_*) - \Psi_{\exp}(x, x_\phi) \right\}^2 \quad , \qquad (2)$$

where $\Psi_{\exp}(x, x_\phi)$ refers to the flux profiles obtained experimentally, $x_\phi = \sqrt{(\max \Phi(x)/\Phi(a)}$ and $\Phi(x)$ is the toroidal flux, as $x_* = 0.2 + 1.1 x_\phi^2 - 0.3 x_\phi^4$.

**Orbit description**

The knowledge of three constants of motion is sufficient for determining the guiding center orbits in a plane spanned by $r$ and the poloidal angle $\chi$. Taking the energy $E$ and the magnetic moment $\mu$ of an ion moving in a field **B** as invariants, and assuming axisymmetry, we derive from the conservation of the toroidal canonical momentum the relation

$$[(x-x_*)\Theta(x-x_*)-(x_i-x_*)\Theta(x_i-x_*)+dh_i\xi_i]^2 - d^2h\left[h-\left(1-\xi_i^2\right)h_i\right] = 0, \quad (3)$$

where $h = A + x\cos\chi$ and $d = mcva/(e\Psi_a')$ identifies the gyroradius of an ion with mass $m$ and velocity $v$ in the poloidal field, and $\xi = \mathbf{v}\cdot\mathbf{B}/(vB)$ is the ion's pitch angle. From Eq.(3) the orbits can be easily found in the poloidal cross section if the initial values marked with the subscript "$i$" are specified at the starting point $(x=x_i, \chi=\chi_i)$ that is chosen most conveniently as the radial maximum $(x_M)$ or minimum $(x_m)$ of an orbit, which occur in the equatorial plane, i.e. $\cos\chi_{i=M,m} = \pm 1$.

Referring to values $A = 3$ and $x_* = 0.2$ typical for recent JET current hole scenarios [7,10,11], and taking $x_i = 0.4$, $\cos\chi_i = -1$, a topological classification of orbits with respect to $\xi_i$ and their extremum radial excursions is illustrated in Fig.1, where the full line exhibits the maximum radial positions of orbits if $x_i \equiv x_m$ and the dotted curve, respectively, represents the location of orbit minima when $x_i \equiv x_M$ is chosen. The range of counter-circulating particles extends from P1 $(\xi_i = -1)$ to P2 $(\xi_i = \xi_{sep}^-)$ lying on the separatrix, beyond which the trapped particle domain $(\xi_{sep}^- \leq \xi_i < 0)$ appears. In Fig.1 an increase of the orbit width $x_M - x_{i=m}$ becomes evident from P1 to P2, where it jumps to a significantly higher value. The corresponding boundary orbit is the so-called fattest banana. The right hand boundary $\xi_i = 0$ defines D-shaped orbits representing the changeover from trapped to co-circulating particles. Moving further from P3a $\rightarrow$ P3b the orbit width will decrease until it vanishes at the stagnation point P3b $(\xi_i = \xi_{stg})$ where $\dot{r} = 0$ and $\dot{\chi} = 0$, i.e. the guiding center will remain at this position. Trajectories with $\xi_m > \xi_{stg}$ cannot be realized, at given $x_m$. As we keep on increasing $\xi_i$ beyond P3b in Fig.1, the minimum and maximum orbit values have interchanged their roles, i.e. the starting points are now maxima, $x_i = x_M$. Note that between P4 and P5 all orbits cross the current hole $(x_* = 0.2)$, whereas from P5 $(\xi_i = \xi_{sep}^+)$ to P6 the orbits are co-going. As apparent from this examination, small perturbations of $x_m$ and $x_M$, especially scattering across the separatrix, can lead to substantial radial displacements of ions.

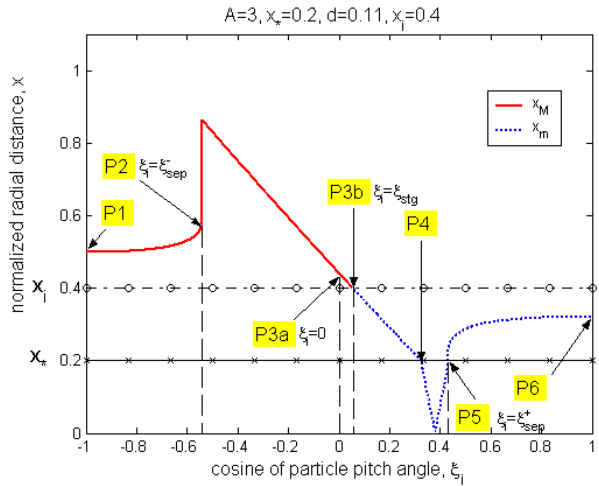

Fig. 1: *Extremum radial positions of guiding center orbits as a function of initial position $x_i$ and initial pitch angle $\xi_i$ for d=0.11.*

## Current hole effect on fast ion confinement domains

Since the plasma particles in tokamaks usually accumulate in the central region near the minor axis, the investigation of the confinement of ion orbits crossing a current hole that is located there, is of evident interest. To prevent from potential double counting of orbits categorized by Fig.1, we turn to a depiction in a plane spanned by $x_M$ and the normalized magnetic moment $\lambda = \mu B(x=0)/E$, which are constants of motion at constant $d \propto v/I$ which, due to a practically steady total plasma current, corresponds with constant ion energy $E$. In Fig. 2 we display the confinement domain of 3.5 *MeV* alphas in a typical current hole JET scenario (A=3, $\gamma = 0.67$, $x_* = 0.2$) [10]. Obviously, the boundary of the confinement domain is formed by $x_M = 1$ and $\lambda$ at the stagnation orbits. To elucidate the locations of the several orbits discussed previously, we arbitrarily take $x_M = 0.7$ and start our inspection in Fig.2a at point Q1 ($\lambda = 0$) associated with a co-circulating orbit with $x_m$ on the left side of the current hole. As $\lambda = (1-\xi^2)h$ increases, i.e. decreasing $\xi$, the orbit size grows, while simultaneously augmenting the orbit width, until it becomes D-shaped (on the left side of the magnetic axis) at Q2 ($\lambda = \lambda_D$), from where the domain of trapped particles is entered. Further reduction of the pitch angle cosine makes the orbits kidney shaped until the fattest banana is realized ($\lambda = \lambda_{sep}$, $\xi = \xi_{sep}^+$). Continuing to higher $\lambda$, the bananas will cross now the current hole while shrinking in size. At point Q3 ($x_m = x_*$, $\xi_m > 0$) we leave the domain of trapped particles that pass the current hole. The next point Q4, a representative of the boundary between trapped and co-circulating ions, indicates the second D-shape orbit possible on the chosen track. Passing the narrow region between Q4 and Q5 ($\lambda = \lambda_{stg}$), not only the size but also the width of orbits is seen to decrease until reduction to the stagnation point. In the confinement domain for counter-going particles in Fig. 2b, where the locus of the separatrix is seen to be equivalent with that of stagnation orbits, only two regions are to be distinguished, the one with orbits outside the current hole and the other with orbits crossing the hole area. The latter region appears considerably smaller than in the case of co-going orbits, where it extends nearly to the wall for the chosen *I*.

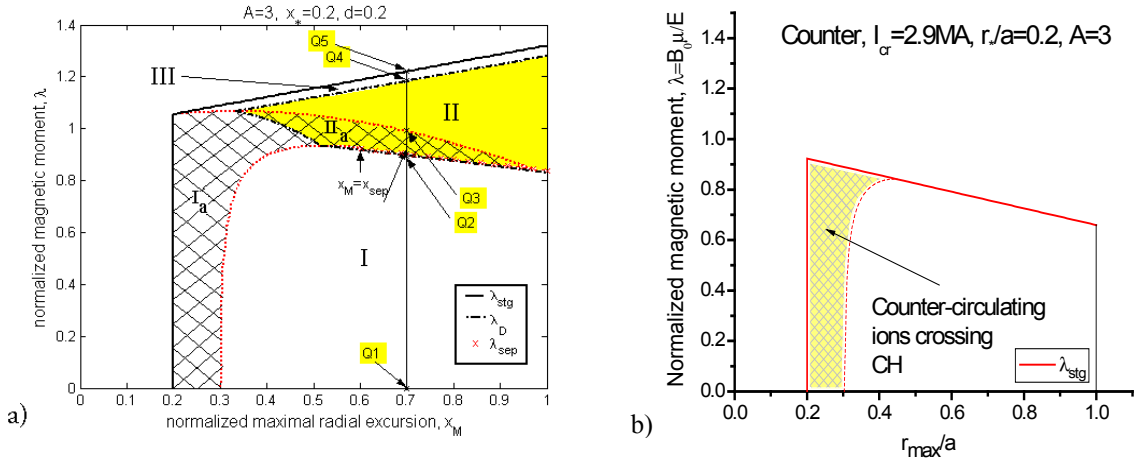

Fig. 2: *Confinement domains for E=3,5MeV alphas for* a) *co-circulating (regions I, Ia and III) and trapped (regions II+IIa) and* b) *counter-circulating orbits in the case of JET-like flux surface shapes (A=3, $\gamma = 0.67$), I=2,9 MA and a small current hole ($x_* = 0.2$). The hatched areas indicate orbits crossing the current hole (CH).*

Comparison of the confinement domains of low/moderate and high energy ions in a current hole tokamak [13] shows that, in the low energy case, there exists only a tiny fraction of trapped particles passing the current hole, whereas, for high energy ions the region of trapped ions crossing the current hole appears substantially larger. Moreover, the latter may not appear as a closed area within $x_M < 1$. Consequently, the portion of trapped fast ions that move through the current hole on orbits with $x_M > 1$ will hit the torus wall. This first orbit loss of fast ions is independent of collision and transport processes, and occurs as well for particles that do not pass the current hole. This confinement deterioration by the presence of a current hole becomes exceptionally strong in the case of relatively low total plasma currents. In order to achieve good fast ion retention in the plasma, the total current has to be $\geq I_{crit}(E = 3,5 MeV, x_*)$, a critical value [10] required to confine the fattest banana orbits crossing the central plasma region. Note that for Fig. 2a $I = I_{crit}$ was chosen enabling the confinement of all trapped fusion alphas crossing the current hole. Equivalently, for a given plasma current $I$ a critical energy can be identified for fast ions of mass $m$ and charge $Z$ as

$$E_{crit}[MeV] \approx \frac{A}{m}\left[Z\gamma\left(1-\sqrt{x}\right)I[MA]\right]^2, \qquad (4)$$

below that all ions crossing the current hole are confined. Therefore, a larger current hole requires an according increase of the plasma current, if the same confinement quality is to be achieved for fast ions. For fusion alphas this CH effect becomes evident from Fig. 3 referring to a larger current hole ($x_* = 0.4$), when compared with Fig. 2. Note that, for each case, the plasma current has been taken equal to the respective $I_{crit}(E_{birth}, x_*)$ in order that all trapped alphas can be retained and contribute to fusion power heating. Quantitatively, as deduced by Figs. 2 and 3, increasing the current hole by doubling the effective hole size of the poloidal flux $\Psi$ from $x_* = 0.2$ to $0.4$ demands for plasma current enhancement by ~50% in order to maintain the same alpha confinement quality.

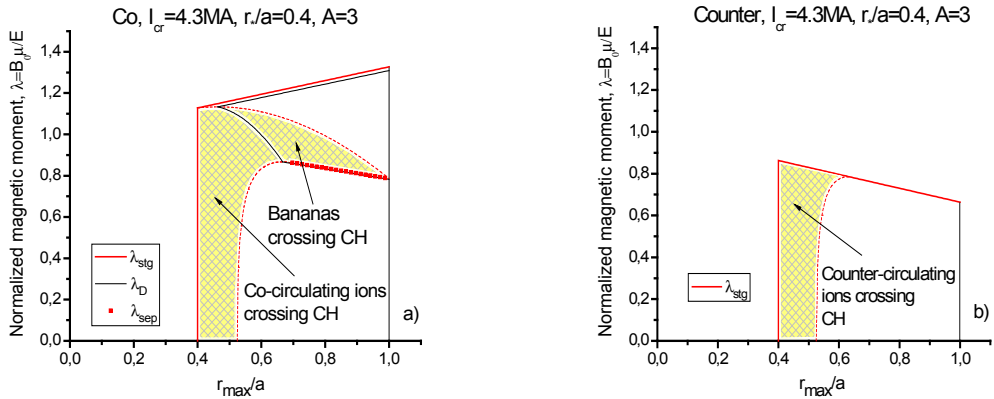

Fig. 3: *Confinement domains of 3,5 MeV alphas for* a) *co-going and* b) *counter-going orbits in the case of JET-like flux surfaces (A=3, $\gamma = 0.67$), I=4,3 MA and a large current hole ($x_* = 0.4$).*

**Fast ion distribution in constants-of-motion space**

An adequate description of the distribution function of high-energetic ions in tokamak plasmas is provided by the bounce averaged drift kinetic equation in the constants-of-motion (COM) Lagrangian space $\mathbf{c} \equiv \{c^1, c^2, c^3\}$ using a Fokker-Planck collision operator,

$$\frac{\partial f}{\partial t} = \frac{1}{\sqrt{g_c}}\frac{\partial}{\partial c^i}\sqrt{g_c}\left(\langle U^i \rangle - \langle D^{ij} \rangle \frac{\partial}{\partial c^j}\right)f + \langle S \rangle, \quad \sqrt{g_c} = \int_{x_m}^{x_M} dx\sqrt{g}, \quad i,j = 1,2,3. \qquad (5)$$

Here $f$ denotes the distribution function, $\sqrt{g}$ is the Jacobian for transformation from Eulerian coordinates $\{\mathbf{x}, \mathbf{v}\}$ to COM Lagrangian space, $U^i$ and $D^{ij}$ are transport coefficients describing convective and diffusive transport in $\{x=r/a, \mathbf{c}\}$ space, $S$ is the fast ion source term, $x_M$ and $x_m$ represent the maximum and minimum values of the normalized radial coordinate along the orbit and $<\ldots>$ indicates bounce averaging in accordance with the rule

$$<X> = \int_{x_m}^{x_M} dx \sqrt{g} X \Bigg/ \int_{x_m}^{x_M} dx \sqrt{g} \equiv \int_{x_m}^{x_M} dx X/\dot{x}(x,\mathbf{c}) \Bigg/ \int_{x_m}^{x_M} dx/\dot{x}(x,\mathbf{c}), \tag{6}$$

where $\dot{x}(x,\mathbf{c})$ is the normalized radial component of the guiding center velocity. In the case of a mono-energetic source term $S=S_0\delta(E-E_0)$ an approximate expression for the initial distribution function of fast ions, which more precisely constitutes a distribution of initial orbits, can be derived from Eq. (5) as

$$f_0 := f(E=E_0, \lambda, x_M) \approx \langle S_0 \rangle \Big/ \langle U^1 \big|_{E_0} \rangle, \tag{7}$$

if $\mathbf{c} \equiv \{E, \lambda, x_M\}$ and $U^1$ denotes the energy slowing down term. We note that for some specific cases the RHS of Eq. (7) allows for an explicit analytical form of the initial ion distribution. Particularly, solutions for stagnation orbits and near stagnation orbits can be derived analytically. Considering a thin tritium beam ($E_0=105$ keV) co-injected in the vicinity of the tokamak mid-plane (on-axis beam, vertical coordinate $Z=Z_{axis}$) [11,14], the beam transforms into initial orbits shown in Fig. 4, while most of the tritons will be seen to execute near stagnation orbits resulting in a density peak there, as displayed in Fig. 5. Accounting for this kind of injection by $S_0 \sim S_{00}\delta(Z-Z_{axis})$ and by

$$\dot{x}(x,\mathbf{c}) \propto Z - Z_{axis} = \pm a\sqrt{x^2 - [h(x,\mathbf{c})-A]^2} = \sqrt{(x_M-x)(x-x_m)}\, G(x,\mathbf{c}) \tag{8}$$

with $G(x_m<x<x_M, \mathbf{c}) \neq 0$, the bounce averaged beam ion source distribution $<S_0>$ in the vicinity of the focusing point ($x_f$ = minimum radial coordinate along the beam trajectory) where ions with near-stagnation orbits are generated, is found as

$$\langle S_0 \rangle \propto \frac{S_{00}}{x_M - x_{stg}(\lambda)} \tag{9}$$

simplifying its maximum location.

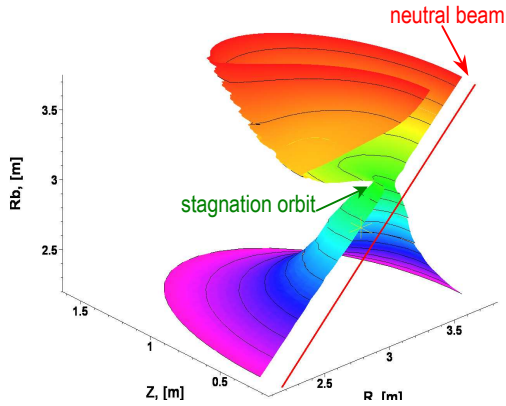
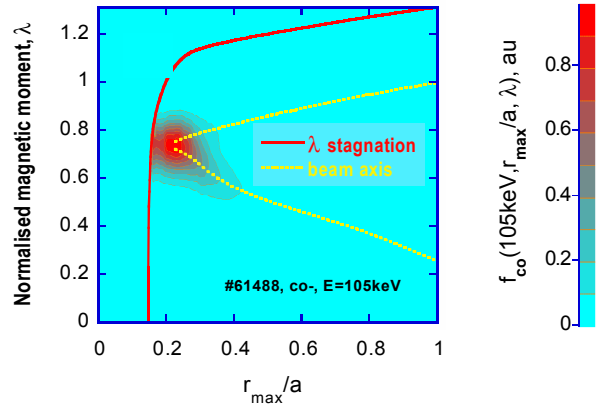

Fig. 4: *Transformation of beam into initial guiding center orbits.*

Fig. 5: *Confinement domain and the beam trace in the $\lambda, r_{max}$-plane for on-axis 105keV co-injected tritium beam ions in a JET current hole plasma (shot #61488) with I/B =1.5MA/3.5T.*

For the 105keV beam triton distribution $f_t$ in the JET CH discharge #61488, numerical Fokker-Planck calculations [11] as well as an analytical derivation using Eq. (9) with $S_{00} \sim \exp[-k^2 a^2 (x_M-x_f)^2/a_b^2]$, corresponding to a Gaussian distribution of beam neutrals perpendicular to the mid-plane with half-width $a_b$=12cm, $x_f$=0.23 and elongation $k$=1.7, lead us to the initial distribution function around the focusing point (at $\lambda=\lambda_f$=0.73 and $x = x_f$=0.23) as depicted in Figs. 6 and 7. Take notice in Fig. 6 of the good agreement between the analytically obtained and the numerical distribution function, thereby confirming the reliability of the approach used. Whereas, in the case of a monotonic current profile, the stagnation orbits are realized close to the magnetic axis, they seem to be shifted outwards by the presence of a current hole.

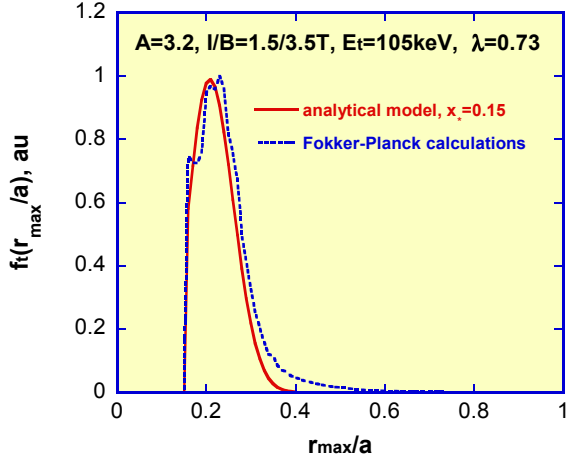 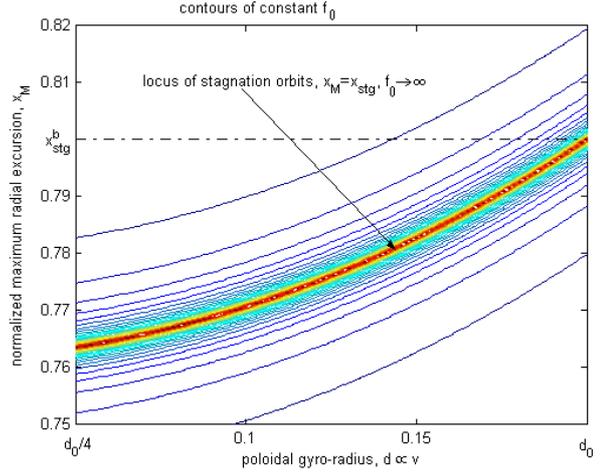

Fig.6: *Distribution function $f_t$ of 105keV beam tritons vs $x_{max}$ in the vicinity of the focusing point in the JET CH discharge #61488.*

Fig. 7: *Contour plot of the stationary distribution function for near stagnation orbits. Here $d(E_0) = d_0$ is the initial poloidal gyroradius.*

Since the poloidal gyroradius is $d \propto v \propto \sqrt{E}$, an important effect of slowing down is the decrease of the maximum radial excursion of orbits as seen from Fig. 7, where the initial NBI triton distribution for stagnation orbits ($f_0(x = x_{stg}) \to \infty$, finite triton density due to $\sqrt{g_c} = 0$) and near stagnation orbits is displayed in the $(d, x_M)$-plane. Also observed can be the broadening of $f_0(E)$ with smaller $x_M$.

**Conclusions**

The model proposed yields manageable analytical expressions describing the entire influence of a current hole on the confinement domains of fast ions as well as on the boundaries separating the regions of different-type orbits in the constants-of-motion space. For specific cases of relevance, i.e. on axis injected beam ions on near stagnation orbits where the maximum density occurs, even an analytical treatment of the bounce averaged Fokker-Planck equation is rendered possible and delivers distribution functions in the COM space in satisfactory agreement with quantitative numerical Fokker-Planck simulations. Hence this suggests the use of the simplified model for further investigations of fast particle behavior in current hole tokamak plasmas.

**Acknowledgement**

This work, supported by the European Communities under the Contract of Association between EURATOM and the Austrian Academy of Sciences, was carried out within the framework of the European Fusion Development Agreement. The views and opinions expressed herein do not necessarily reflect those of the European Commission.